\documentclass[a4paper,conference]{IEEEtran}

\usepackage{cite}
\usepackage{amsmath,amssymb,amsfonts}
\usepackage{graphicx}
\usepackage{bm}
\usepackage{booktabs}
\usepackage{multirow}
\usepackage{url}
\usepackage{setspace}
\usepackage{comment}

\begin{document}
\small
\begin{spacing}{1}

\title{Impact of Expert-Following Strategies\\ in Financial Asset Recommendation}

% TODO: 著者情報を記入
\author{\IEEEauthorblockN{Ryuki Unno$^{\dagger}$, Koshi Watanabe$^{\dagger\dagger}$, Keigo Sakurai$^{\dagger\dagger\dagger}$, Keisuke Maeda$^{\dagger\dagger\dagger}$, Takahiro Ogawa$^{\dagger\dagger\dagger}$ and Miki Haseyama$^{\dagger\dagger\dagger}$}
\IEEEauthorblockA{
$^{\dagger}$ School of Engineering, Hokkaido University, Japan\\
$^{\dagger\dagger}$ Graduate School of Information Science and Technology, Hokkaido University, Japan\\
$^{\dagger\dagger\dagger}$ Faculty of Information Science and Technology, Hokkaido University, Japan\\
E-mail: \{unno, koshi, sakurai, maeda, ogawa, mhaseyama\}@lmd.ist.hokudai.ac.jp}
\vspace{-5mm}
}

\maketitle
\IEEEpeerreviewmaketitle

% =========================================================
\begin{abstract}
Financial institutions hold rich transaction histories, yet delivering recommendations that simultaneously maximize investment returns and ensure preference alignment remains a significant challenge.
Existing approaches, namely return-based and preference-based strategies, each optimize a single objective, resulting in a fundamental trade-off between profitability (ROI) and relevance (nDCG).
In this paper, we propose the \textit{Expert-Following Strategies}: a framework that identifies top-performing investors based on their historical ROI and recommends the assets they purchased, scored by ROI-weighted purchase frequency.
Our experiments using real-world transaction histories show that our strategy achieves statistically significant improvement over the market-average baseline in both ROI and nDCG simultaneously across all four thresholds.
\end{abstract}

% =========================================================
\section{Introduction}
Investment advisory services are foundational to financial institutions, aiming to facilitate informed investment decisions by matching investors with suitable assets.
To date, such advice has relied on the subjective domain expertise of financial advisors.
In recent years, data-driven approaches using transaction histories have been attracting attention.
With the development of data-driven investment advice, the automation of individual-dependent tasks is expected.

\textit{Financial Asset Recommendation} (FAR), which applies a recommender system based on purchase histories, is gaining significant traction.
FAR typically applies general recommendation algorithms such as collaborative filtering, where latent preferences are modeled by representing past investor purchase records as graph structures or similar formats and learning the embeddings of investors and financial assets.
This enables identifying unpurchased assets that align with latent preferences, thereby personalizing the investment experience.

However, unlike conventional recommendation tasks, FAR requires the consideration of not only preference (normalized Discounted Cumulative Gain (nDCG)) but also actual capital gains (Return on Investment (ROI))~\cite{sakurai2026}.
On the investors' side, preferences inherently include expectations for investment returns; on the service provider side, generating practical benefits is essential to encourage continuous use.
While simultaneous optimization of ROI and nDCG is required from both perspectives, a trade-off exists between the two in the existing FAR framework~\cite{sanzcruzado2024}.
If only investment returns are emphasized, the investor's preferences will be ignored, and if only preferences are emphasized, returns will remain at the market average level.
In fact, Sanz-Cruzado et al.~\cite{sanzcruzado2024} confirmed this trade-off in 11~algorithms and pointed out that there is no method that is simultaneously superior in both ROI and nDCG.
This limitation necessitates incorporating behavioral signals to jointly optimize these competing criteria.

To address this problem, this paper proposes \textit{Expert-Following Strategies} in FAR, a new approach designed to strike the optimal balance between individual preferences and actual capital gains.
In a previous study~\cite{coval2021}, it was reported that high-performing investors in financial markets consistently achieve high rates of return.
We assume that they reflect both a strategy to maximize profits (ROI) and realistic purchasing behavior that can also match the preferences of general investors (nDCG).
Their transaction patterns contain both profitability and behavioral signals, and it is expected that expert-based recommendation will be a potential optimal balance in the space of ROI and nDCG.
In this paper, we propose a framework that identifies the top $X$\% of investors in past transactions and recommends assets based on their transaction histories, which does not require complex model learning.
As a result of the experiments, it was confirmed that this approach can simultaneously achieve high ROI and high nDCG compared to general baselines.

% =========================================================
\section{Method}

\subsection{Problem Formulation}
We represent data as triplets $(u,\,i,\,t)$, where an investor $u$ purchased asset $i$ on day $t$, and
$p_{i,t}$ denotes the closing price of an asset $i$ on day $t$.
For evaluation, purchase histories are considered, and all positions are assumed to be held until the final day of the evaluation window.
Given a decision time $t$, the task is to recommend a ranked list of unseen assets for investors.
All methods recommend the top-$k$ assets, assigning them equal weight $w_i=1/k$.
The performance is evaluated over a subsequent window $(t, t + \Delta t]$.
This window $\Delta t$ serves as both the holding period for ROI and the observation period for nDCG. Following \cite{sanzcruzado2024}, we set $k=10$ and $\Delta t=6$~months.

\subsection{Baselines}

We evaluate approaches from opposite sides of the ROI--nDCG trade-off, plus one reference.
\begin{itemize}
  \item \textbf{Return-based Strategies:}
    \begin{itemize}
      \item \textbf{Sharpe Ratio Optimization (SR~Opt.)}

        This method optimizes the weights by maximizing the annualized Sharpe Ratio
        \begin{equation*}
          \max_{\bm{w}}\;\sqrt{252}\,\frac{\bm{w}^{\!\top}\bm{\mu}}{\sqrt{\bm{w}^{\!\top}\bm{\Sigma}\bm{w}}},
        \end{equation*}
        where $\bm{\mu}$ is the mean daily return vector and $\bm{\Sigma}$ is the Ledoit-Wolf shrinkage covariance matrix~\cite{ledoit2004}.
      \item \textbf{Random Forest (RF)}
        
        This method estimates the ROI using 11~technical indicators computed from $p_{i,t}$.
        It splits the training period into two halves and selects top-$k$ assets by predicted ROI~\cite{breiman2001}.
    \end{itemize}
  \item \textbf{Preference-based Strategies:}
    \begin{itemize}
      \item \textbf{Popularity (Pop.)}
        
        This method selects the top-$k$ assets by purchase transaction count in training period~\cite{sanzcruzado2024}.
    \end{itemize}
  \item \textbf{Baseline:}
    \begin{itemize}
      \item \textbf{Market~Average (Market~Avg.)}
        
        This method selects all $N$ assets at equal weights.
    \end{itemize}
\end{itemize}
Standard CF methods are excluded because \cite{sanzcruzado2024} shows they do not surpass Popularity on FAR-Trans.

\subsection{Expert-Following Strategies}
Unlike single-objective baselines, we propose \textit{Expert-Following Strategies} to jointly maximize ROI and nDCG. This framework leverages the insight that high-performing investors' transaction histories inherently link high returns to realistic purchase behavior.

\textbf{Expert identification.}
For investor $u$, we compute their ROI over the training period $[t_0,\,t]$ as the portfolio return:
\begin{equation}
  \mathrm{ROI}_u(t_0, t)
    = \frac{V_u(t) - V_u(t_0)}{V_u(t_0)},
  \label{eq:roi}
\end{equation}
where $V_u(t) = \sum_i w_{u,i}\,p_{i,t}$ is the market value of $u$'s portfolio at $t$, and $w_{u,i}$ is the net number of shares of asset $i$ currently held by investor $u$ (purchases minus sales, clipped to zero).

\textbf{Asset scoring.}
To select the top-$k$ assets for recommendation, we assign a recommendation score $S_i$ for each asset $i$.
$S_i$ is scored by ROI-weighted purchase frequency among experts:
\begin{equation}
  S_i
    = \sum_{u \in \mathcal{E}_X}
      \max\!\bigl(\mathrm{ROI}_u(t_0, t),\,0\bigr) \cdot C_{u,i},
  \label{eq:score}
\end{equation}
where $C_{u,i}$ is the purchase count of asset $i$ by expert $u$ during $[t_0,\,t]$, and the top-$X\%$ of investors by $\mathrm{ROI}_u(t_0, t)$ form the expert set $\mathcal{E}_X$.
This formulation combines two signals:
purchase frequency captures collective expert preference, while ROI weighting ensures top performers exert greater influence, distinguishing our approach from simple filtered popularity.
The top-$k$ assets by $S_i$ are recommended.
We evaluate four thresholds $X \in \{5,\,10,\,15,\,20\}$ to assess how the strictness of expert selection affects the ROI--nDCG balance, where smaller $X$ yields a stricter, more concentrated expert group.
% =========================================================
\section{Experiments}

\begin{table}[!t]
\centering
\caption{Multi-instance evaluation results ($n=10$).
  \textbf{Bold}: best per metric (Reference rows excluded).
  $^\star$: significant improvement vs.\ market average (both paired $t$-test and Wilcoxon signed-rank test, Bonferroni-corrected $\alpha'\!=\!0.05/7\!\approx\!0.0071$).
}
\label{tab:main}
\renewcommand{\arraystretch}{1.12}
\setlength{\tabcolsep}{3.5pt}
\footnotesize
\begin{tabular}{@{}llccc@{}}
\toprule
\textbf{Category} & \textbf{Method} & \textbf{ROI@10} & \textbf{SR@10} & \textbf{nDCG@10} \\
\midrule
\multirow{4}{*}{\shortstack[l]{Expert-\\Following}}
  & top-5\%        & \textbf{0.246}$^\star$ & 1.185 & 0.176$^\star$ \\
  & top-10\%       & 0.133$^\star$          & 0.627 & 0.181$^\star$ \\
  & top-15\%       & 0.131$^\star$          & 0.590 & 0.185$^\star$ \\
  & top-20\%       & 0.131$^\star$          & 0.583 & 0.185$^\star$ \\
\midrule
Pref.-based & Pop.     & 0.100 & 0.350 & \textbf{0.201}$^\star$ \\
\midrule
\multirow{2}{*}{\shortstack[l]{Return-\\based}}
  & RF             & 0.218$^\star$ & \textbf{1.895}         & 0.008 \\
  & SR~Opt.        & 0.095         & 1.480                  & 0.002 \\
\midrule
Reference & Market~Avg.    & 0.080 & 3.325 & 0.007 \\
\bottomrule
\end{tabular}
\end{table}

\begin{figure}[!t]
\centering
\includegraphics[width=\columnwidth]{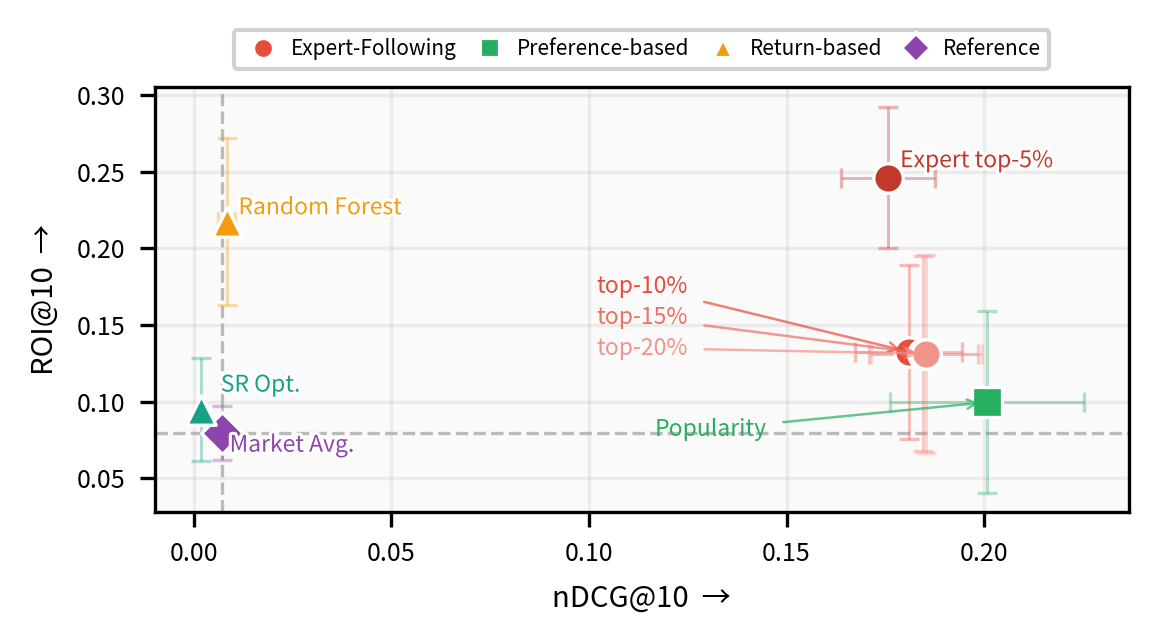}
\caption{ROI@10 vs.\ nDCG@10 (mean $\pm$ std, $n\!=\!10$ instances).
  Expert-Following (top-5\%) uniquely occupies the high-ROI/high-nDCG region; dashed lines mark the Market~Average baseline.}
\label{fig:scatter}
\end{figure}

\subsection{Settings}

To ensure robustness and reduce evaluation variance, we adopt a multi-instance walk-forward validation framework on FAR-Trans (Jan~2018 -- Nov~2022).
Reserving the first 18~months as an initial training buffer, we generate 61~checkpoints at two-week intervals (Jul~2019 -- Apr~2022).
At each checkpoint $t$, a method trains on all transactions up to $t$ and evaluates performance over the following $\Delta t$.
We construct 10~sliding-window instances, each with training checkpoints $[j,\,j+29]$ (${\approx}$14~months) and evaluation checkpoints $[j+30,\,j+49]$ (20~time points, ${\approx}$10~months), averaging results across instances.
We assume zero transaction costs and slippage to focus on the intrinsic predictive power of the strategies.
We report ROI@10 (mean portfolio return over $\Delta t$), SR@10~\cite{sharpe1966} (annualized Sharpe~Ratio), and nDCG@10~\cite{jarvelin2002} (purchases within $\Delta t$ as relevant items).
Each of the seven non-reference methods is compared against the market-average baseline using a paired $t$-test and Wilcoxon signed-rank test with Bonferroni correction ($\alpha'=0.05/7\approx0.0071$).

\subsection{Results}
Table~\ref{tab:main} confirms the ROI--nDCG trade-off: return-based strategies (e.g., RF) achieve competitive ROI~(0.218) but near-zero nDCG~(0.008), whereas preference-based strategies (e.g., Pop.) attain the highest nDCG~(0.201) but no significant ROI gain over the market average.
The Expert-Following Strategies effectively optimize this trade-off.
As shown in Fig.~\ref{fig:scatter}, all expert variants achieve a superior balance compared to the baselines; for instance, top-5\% achieves the highest ROI~(0.246) while maintaining a significant nDCG~(0.176).
These results empirically validate that expert investors inherently exhibit behavioral patterns leading to both capital gains and preference alignment, confirming expert behavior as a robust foundation for financial asset recommendation.

% =========================================================
\section{Conclusion}
We proposed \textit{Expert-Following Strategies}, a simple FAR framework that identifies top-ROI investors and recommends the assets they purchased, scored by ROI-weighted purchase frequency.
Evaluated on FAR-Trans with a multi-instance walk-forward validation framework and rigorous significance testing, Expert-Following variants are the only methods achieving statistically significant improvement over the market-average baseline in both ROI and nDCG simultaneously: directly resolving the trade-off between ROI and nDCG.
Limitations include assuming zero transaction costs, which may overestimate ROI for high-turnover strategies (e.g., top-5\%).
Future work will explore realistic transaction fees and graph-based recommendation models.

% =========================================================
\end{spacing}

\vspace{-0pt}
\small
\scriptsize
\bibliographystyle{IEEEtran}
\bibliography{ref.bib}

\end{document}